\documentclass[11pt,superscriptaddress]{revtex4-2}   
\pdfoutput=1
\usepackage{amsmath}
\usepackage{graphicx}
\usepackage{natbib}
\usepackage{mathrsfs}
\usepackage{ulem}
\usepackage{hyperref}
\usepackage[T1]{fontenc}
\usepackage[letterpaper,textwidth=7in,top=.75in,bottom=.75in]{geometry}
\usepackage{color}
\usepackage[dvipsnames]{xcolor}

\usepackage{soul}

\begin{document}
\title{Sensing electrochemical signals using a nitrogen-vacancy center in diamond}

 \author{Hossein T. Dinani\footnote{htdinani@gmail.com}}
  \address{Facultad de F\'isica, Pontificia Universidad Cat\'olica de Chile, Santiago 7820436, Chile}
      \address{
     Centro de Investigaci\'{o}n DAiTA Lab, Facultad de Estudios Interdisciplinarios,\\ Universidad Mayor, 	Santiago, Chile}

 \author{Enrique Mu\~noz\footnote{munozt@fis.puc.cl}}
 \address{Facultad de F\'isica, Pontificia Universidad Cat\'olica de Chile, Santiago 7820436, Chile}
    \address{Research Centre for Nanotechnology and Advanced Materials, Pontificia Universidad Cat\'olica de Chile, Santiago, Chile}

 \author{Jeronimo R. Maze}
  \address{Facultad de F\'isica, Pontificia Universidad Cat\'olica de Chile, Santiago 7820436, Chile}
    \address{Research Centre for Nanotechnology and Advanced Materials, Pontificia Universidad Cat\'olica de Chile, Santiago, Chile}

\date{\today}

\begin{abstract}
Chemical sensors with high sensitivity that can be used in extreme conditions and can be miniaturized are of high interest in science and industry. The Nitrogen-vacancy (NV) center in diamond is an ideal candidate as a nanosensor due to the long coherence time of its electron spin and its optical accessibility. In this theoretical work, we propose to use an NV center to detect electrochemical signals emerging from an electrolyte solution, thus obtaining a concentration sensor. For this purpose, we propose to use the inhomogeneous dephasing rate of the electron spin of the NV center  ($1/T^{\star}_2$)  as a signal. We show that for a range of mean ionic concentrations in the bulk of the electrolyte solution, the electric field fluctuations produced by the diffusional fluctuations in the local concentration of ions, result in dephasing rates which can be inferred from free induction decay measurements. Moreover, we show that for a range of concentrations, the electric field generated at the position of the NV center can also be used to estimate the concentration of ions.
\end{abstract}
\maketitle

\section{Introduction}
Accurate detection of ionic concentrations in an electrolyte solution is of fundamental interest for process control and optimization in the chemical, farmaceutical and food industry. It is also important in environmental applications as a means to detect pollutant levels, as well as in fundamental studies in chemistry and biology. Examples are measuring concentration of copper ions in a copper refinery, and measuring pH of a solution which is defined as negative logarithm of concentration of hydrogen ions. In copper electro-refining processes, electrolysis is used to increase the purity of copper in combination with an electrolyte ${\rm Cu}^{2+}/{\rm SO}_4^{2-}$ solution, and monitoring ionic concentration is required to control the purity of copper with high precision.

Carbon based materials have been used widely as sensors\cite{Anjum, Romana, Gattia17,Gattia19}. Among them, the negatively charged nitrogen-vacancy (NV) center in diamond has attracted much attention. Diamond is a bio-compatible crystal resilient to extreme conditions. Moreover, the electron spin of the NV center can be prepared and read out through optical excitation \cite{Doherty_PhysRep}. The zero field splitting between the spin triplet sub-levels in the ground state of the NV center is sensitive to various physical parameters such as crystal strain \cite{Doherty}, temperature \cite{Kucsko}, electric \cite{Dolde} and magnetic field \cite{Rondin, Maze, Bonato, HTD}. The charge state of the NV center can also be used as a sensor.  In Refs.~\cite{Newell, Rendler} the effect of pH of an electrolyte solution on the charge state of the NV center in a bulk diamond, and in a functionalized nanodiamond have been investigated.

In addition, the coherence time of the electron spin of the NV center is sensitive to its surrounding environment. The decoherence time of the NV center is used to estimate fluctuating magnetic fields from paramagnetic impurities \cite{Luan}, liquid's diffusion coefficient \cite{Staudacher}, drift velocity inside a microfluid channel \cite{Cohen}, and concentration of ions in a cell membrane \cite{Hall}. In Ref.~\cite{Fujiwara} the decoherence time of the NV center is monitored during a change in the pH of the electrolyte solution.

Here, we consider an electrolyte-diamond interface as shown in Fig.~\ref{fig:geometry}. We propose to use the dephasing rate of the electron spin of an NV center in diamond in order to estimate the concentration of ions in the electrolyte solution. The rationale behind this choice, as we will explain in detail in the following sections, is that the diffusional fluctuations in the local concentration of ions in the electrolyte solution generate an electrical noise signal, that results in an additional dephasing rate, $1/T^\star_2$, that competes with the intrinsic dephasing rate of the NV center. 
As we will show, for a range of mean ionic concentrations in the bulk of electrolyte, $c_b>0.04$ mol/m$^3$, the induced $1/T^\star_2$ is larger than 10 kHz, and for $c_b>100$ mol/m$^3$ is larger than 300 kHz. Therefore, depending on the intrinsic $1/T^\star_2$ of the electron spin of the NV center, the range of $c_b$ that can be estimated with this method can be larger than 0.04 mol/m$^3$. We also show that for a lower range of ion concentrations, $c_b<0.1$ mol/m$^3$, the gradient of the electric field generated at the position of the NV center is large enough that may be used to estimate the concentration of ions.

\section{Results and discussion}

\subsection{Electrolyte solution}
In this section, we present a brief analysis of the basic physico-chemical mechanisms governing diffusion of ionic chemical species in a liquid electrolyte solution, which is displayed as the sample to be analyzed in the proposed experimental setup (Fig.~\ref{fig:geometry}). Since the ionic species and their concentration profiles determine a local charge density, by solving the Poisson equation we obtain the electric potential and the electric field inside the electrolyte solution. By a further statistical analysis, we then obtain mathematical expressions for the fluctuations of the electric field inside the electrolyte solution.
\subsubsection{Diffusion of ionic species in electrolyte solutions}\label{sec_diffusion}

\begin{figure}[b!]
\centering
\includegraphics[width=0.55\textwidth]{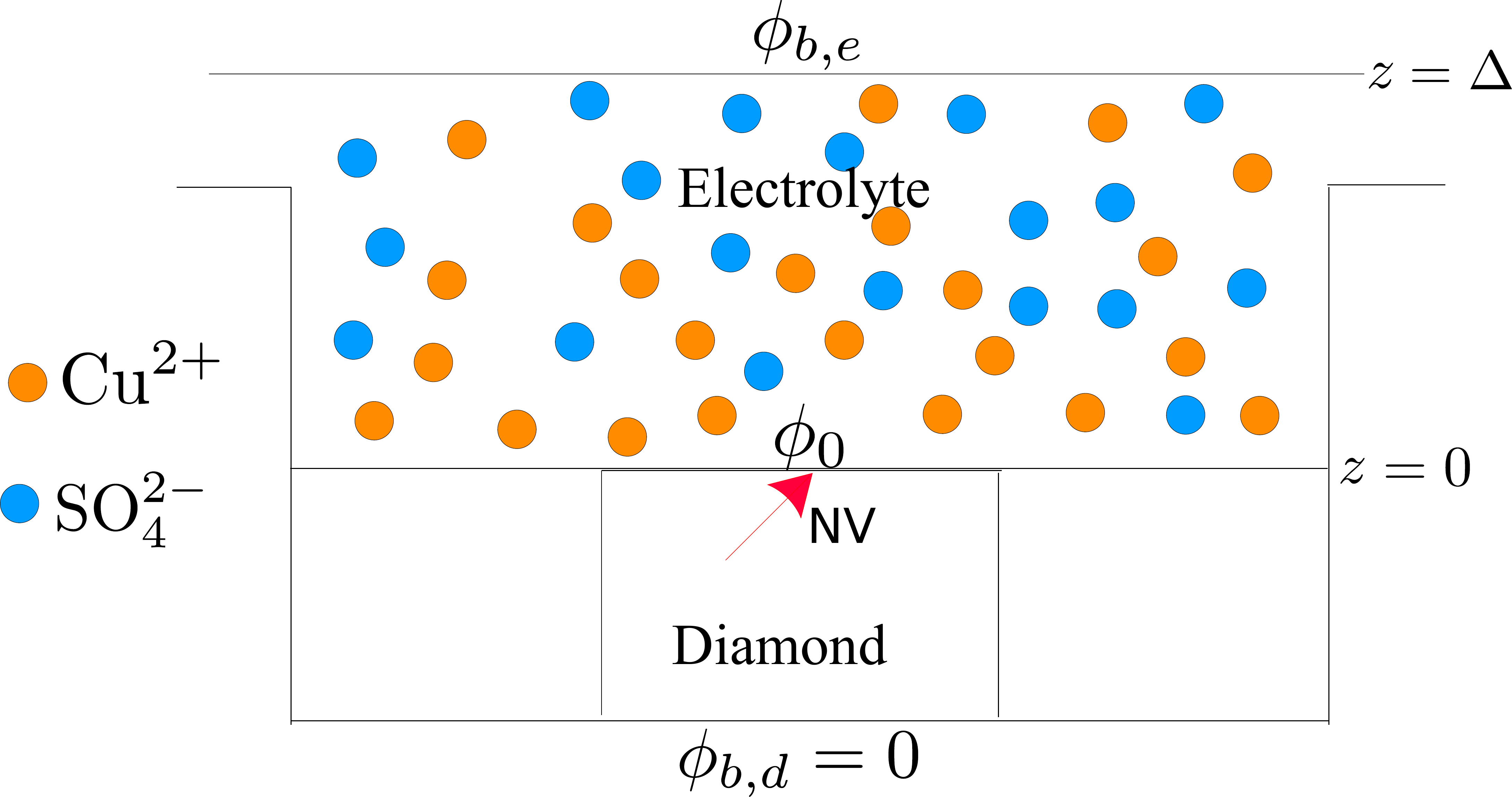}
	\caption{The proposed experimental setup which contains a ${\rm Cu}^{2+}/{\rm SO}_4^{2-}$ electrolyte solution in contact with the surface of a bulk diamond crystal with an NV center (red arrow). The upper surface of the diamond is taken to be at $z=0$ with potential $\phi_0$, while $z=\Delta$ is far into the bulk of the electrolyte solution with potential $\phi_{b,e}$. The potential in the opposite surface of the diamond is shown with $\phi_{b,d}$ taken to be zero in our simulations.}
\label{fig:geometry}
\end{figure}

A typical model for the flux $\mathbf{N}_s$ (mol m$^{-2}$ s$^{-1}$) of ionic species in a liquid electrolyte solution, taking into account the effect of concentration gradient, Fick's law, and the effect of electric field, is given by \cite{Kirby}
\begin{equation}
 \mathbf{N_s}= -\mathcal{D}({\nabla c_s + \frac{z_s c_s \mathcal{F}}{RT}\nabla \phi}).
\end{equation}
In this equation, known as Nernst-Planck equation, $\mathcal{D}$ is the diffusion constant, $z_s$ is the valence of each ion species in the electrolyte solution, and $c_s$ is the concentration of each species. The Faraday constant, $\mathcal{F}=96485.3365$~C/mol, is a measure of charge per mole of substance and is the product of Avogadro's number and the charge of an electron,  $R=8.314$ J/(mol K) is the universal constant of gases, $T$ is temperature in K, and $\phi$ is the electric potential in Volt. By requiring mass balance, we obtain the following linear partial differential equation \cite{Kirby}
\begin{equation}
\label{eq:nernst-planck}
\frac{\partial c_s}{\partial t}+ \nabla \cdot  \mathbf{N_s}=0.
\end{equation}

The above equation accounts for how the diffusion process involves electrical forces on each ionic species in an electrolyte solution. To account for the electric field response due to the presence of the ionic species we consider the Poisson equation
\begin{equation}
\nabla^2 \phi = -\frac{\rho}{\epsilon_e}.
\end{equation}
Here $\epsilon_e$ is the electric permittivity of the electrolyte solution (approximately the same as in water for concentrations below $10^3 ({\rm mol/m^3})$ \cite{Gavish_016}) and $\rho(\mathbf{x})$ is the local charge density determined by the ionic concentrations
\begin{equation}
\rho= \sum\limits_{s=\pm} {\mathcal{F} z_s c_s}.
\end{equation}
 
Due to the small thickness of the fluid layer of the sample as compared to its transverse dimensions, we assume that the system presents concentration gradient and potential gradient only along the direction $z$ normal to the interface (see Fig.~\ref{fig:geometry}). This approximation is equivalent to consider an infinitely large interface surface between the liquid solution and the diamond crystal. Therefore, we can write $\mathbf{N}_{s} = \hat{z}N_{s}$ with the boundary condition $N_{s}|_{z=0}=0$, that expresses the fact that the ionic species in the liquid solution cannot penetrate into the solid crystal. As a result, in the steady state, i.e., ${\partial c_s}/{\partial t} = 0$, from Eq.~(\ref{eq:nernst-planck}) we obtain $\mathbf{N}_s=0$ for $s = \pm$. With the previous considerations, the simplified system of coupled, non-linear differential equations to be solved is given by
\begin{eqnarray}
&&\frac{\partial c^{}_s}{\partial z}+\frac{ z_s\mathcal{F}}{RT}c^{}_s(z)\frac{\partial \phi^{}}{\partial z} =0 \label{eq_dcs},\\
&&\frac{\partial^2  \phi^{}}{\partial z^2} = -\frac{1}{\epsilon_e}\sum_{s= \pm} z_s \mathcal{F} c^{}_{s}(z).
\label{eq:concentration-diff-zero3}
\end{eqnarray}

Taking the integral of Eq.~\eqref{eq_dcs} from $z$ to $\Delta$, with $\Delta$ being the distance from the interface far into the bulk of the solution, we obtain 
\begin{equation}\label{eq_cb}
c^{}_s(z)=c_{b,s}\exp\left(-{\frac{z_s\mathcal{F}}{RT}\left(\phi(z)-\phi(\Delta)\right)}\right),
\end{equation}
where $c_{b,s}$ is the bulk concentration of each species. Taking into account that the electrolyte should be electrically neutral in the bulk, due to charge conservation, we obtain the electric potential as (see Appendix \ref{appendixA})
\begin{equation}\label{eq_phie}
\phi_{}(z) - \phi(\Delta)= \frac{2 R T}{z_s \mathcal{F}}\ln\left[\frac{1+\tanh(\frac{z_s \mathcal{F} V_0}{4 R T})\exp(-\kappa z )}{1-\tanh(\frac{z_s \mathcal{F} V_0}{4 R T})\exp(-\kappa z )}\right].
\end{equation}
Here, $V_0=\phi(0)-\phi(\Delta)$, where $\phi(\Delta)$ is the potential in the bulk of the electrolyte (see Fig.~\ref{fig:geometry}). We have defined $\kappa$, the inverse screening length, as
\begin{equation}\label{eq_kappa}
\kappa^2  = \frac{2 z_s^2\mathcal{F}^2 c_b}{RT\epsilon_e}.
\end{equation}
Note that, Eq.~\eqref{eq_phie} is valid for $\kappa \Delta \gg 1$ (see Appendix \ref{appendixA}). From this potential we can calculate the electric field, which on the surface of the diamond inside the electrolyte is given by
\begin{equation}\label{eq_Ee}
E_{eq}(z=0^+)=\frac{2\kappa R T}{z_s \mathcal{F}}\sinh\left(\frac{z_s\mathcal{F}V_0}{2 RT}\right).
\end{equation}
Here, the subscript $eq$ emphasizes that this corresponds to the
local electric field in thermal equilibrium.

\subsubsection{Electric field fluctuations at the interface between the solution and the diamond} 
Due to thermal noise, the concentration of ions will have small fluctuations around its equilibrium value, i.e., $\delta c(z,t) = c(z,t) - c_{ eq}(z)$. These fluctuations in the concentration will induce fluctuations in the electric field, i.e., $\delta E(z,t) = E(z,t) - E_{ eq}(z)$. The electric field fluctuations are directly linked to the fluctuations in concentration through the Poisson equation
\begin{eqnarray}
\frac{\partial}{\partial z}\delta E(z,t)= -\frac{\partial^2}{\partial z^2}\delta\phi(z,t)  = \frac{\delta \rho(z,t)}{\epsilon_e}
= \frac{\mathcal{F}}{\epsilon_e}\sum_{s = \pm}z_s\delta c_s(z,t).
\end{eqnarray}
We can integrate this equation over $z$ in the range $[z, \Delta]$. For $\kappa \Delta \gg 1$ we have $\delta E(z=\Delta,t) \sim 0$. On the other hand, as is shown in Appendix \ref{AppendixB}, the correlation of the fluctuations of concentration is given by
\begin{eqnarray}
\langle \delta c_s(z_1,t) \delta c_{s'}(z_2,0)\rangle = \delta_{s,s'} \frac{c_s^{eq}(z_1)}{N_A A\left(4 \pi \mathcal{D}_s\, t \right)^{1/2}}\exp\left[-{\frac{(z_1- z_2)^2}{4\mathcal{D}_s t}  }\right]\Theta(t).
\end{eqnarray}

Therefore, the correlation of the fluctuations of the electric field at the surface of the diamond from the liquid solution side, i.e., at $z=0^{+}$ is obtained as
\begin{eqnarray}\label{correlator_Ee}
\langle \delta E(0,t) \delta E(0,0) \rangle =   \frac{\mathcal{F}^2}{N_A A \epsilon^2_e}\sum_{s = \pm}z_s^2  \int_{0}^{\Delta}dv c_s^{eq}(v) \frac{{\rm Erf}\left[\frac{\Delta - v}{\sqrt{4 \mathcal{D}_s t}} \right]  - {\rm Erf}\left[\frac{-v}{\sqrt{4 \mathcal{D}_s t}}\right] }{2}.
\end{eqnarray}
\begin{figure}[b!]
\centering
	\includegraphics[width=0.65\textwidth]{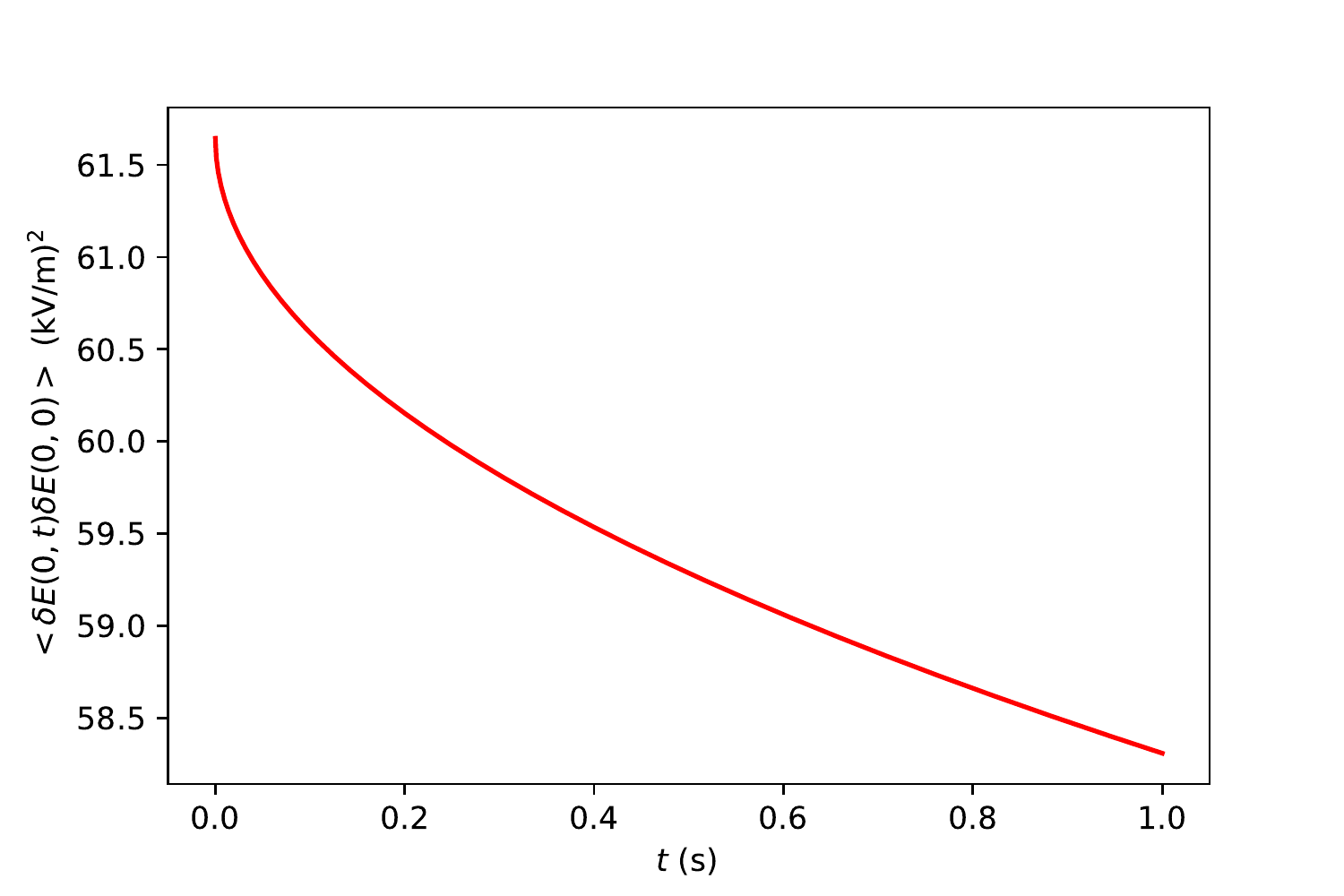}
	\caption{Correlation of the fluctuations of the electric field given in Eq.~\eqref{correlator_Ee_simplified} versus time. We have taken $A=4$ mm$^2$, $\Delta=1$mm, ${\rm c_b=1~mol/m^3}$, $z_s=2$, $\mathcal{D}_+=\mathcal{D}_-=2.3 \times 10^{-9}~ {\rm m^2/s}$.}
\label{fig_deltaE}
\end{figure}
Using the equilibrium concentration given in Eq.~\eqref{eq_cb}, in the limiting case $\kappa\Delta \gg 1$, the electric field fluctuations at the surface of diamond from the liquid solution side, i.e. at $z = 0^{+}$, can be simplified to \begin{eqnarray}\label{correlator_Ee_simplified}
\langle \delta E(0,t) \delta E(0,0) \rangle =   \frac{\mathcal{F}^2 \Delta c_b}{N_A A \epsilon^2_e}\sum_{s = \pm}{ z^2_s \left\{{\rm Erf}\left[\frac{\Delta}{\sqrt{4 \mathcal{D}_s t}} \right] - \frac{1}{\Delta}\sqrt{\frac{4 \mathcal{D}_s t}{\pi}}\left(1-\exp\left({-\frac{\Delta^2}{4 \mathcal{D}_s t}}\right)\right) \right\}}.
\end{eqnarray}
It is clear that the correlation of the fluctuations is directly proportional to the concentration in the bulk $c_b$. Figure \ref{fig_deltaE} shows the correlation of the fluctuations of the electric field on the surface of the diamond as a function of time $t$ for a fixed value of $c_b=1$ mol/m$^3$.

In the following section, we obtain the fluctuations of the electric field at the position of the NV center in diamond. For doing so, we first obtain the electric potential and the electric field inside the diamond.

\subsection{Diamond and the NV center}
The electron spin of the NV center in diamond is sensitive to the electric field and to its fluctuations. The fluctuations of the electric field result in dephasing of the NV electron spin. In this section, we consider an NV center in a bulk diamond in contact with the electrolyte, as depicted in Fig.~\ref{fig:geometry}. The electric field inside the diamond is screened by the dielectric response of the diamond and ionization of donors and acceptors inside the diamond \cite{Broadway}. We first find the electric field and its fluctuations, resulting from the concentration of ions in the electrolyte and their fluctuations, at the position of the NV center. We then find the dephasing rate of the electron spin of the NV center as a result of the fluctuations of the electric field. Finally, we show that for a range of $c_b$, the gradient of the electric field induced at the position of the NV center is large enough that can be resolved by a Ramsey measurement.

\subsubsection{Potential and electric field inside the diamond}
As is shown in Fig.~\ref{fig:geometry}, we consider a bulk diamond in contact with the electrolyte with its interface at $z=0$.
\begin{figure}[b]
\centering
\includegraphics[scale=0.65]{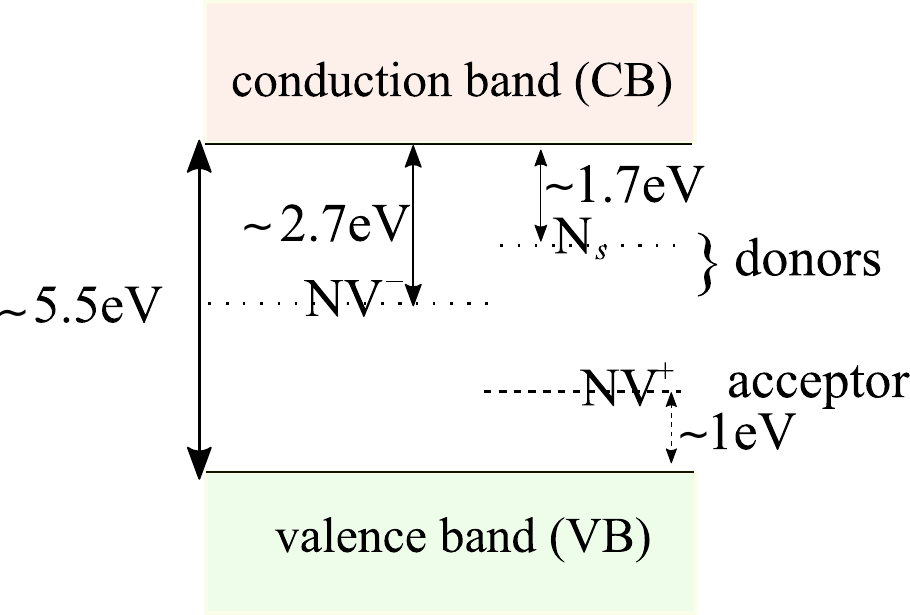}\label{fig_bandgap}
\caption{Bandgap of diamond and the energy level of the defects. Substitutional nitrogen ${\rm N_s}$ and the negatively charged NV, NV$^-$, act as donors, while the positively charged NV, NV$^+$, acts as an acceptor.}
\vspace{0.5cm}
\end{figure}
We assume that the bulk diamond is implanted with nitrogen ions, forming mainly substitutional nitrogen, ${\rm N_s}$, ($\approx 96\%$) and NV defects ($\approx 4\%$), and that the concentration of vacancy defects is negligible. The substitutional nitrogen and the negatively charged NV center, NV$^-$, act as donors with ionization energies $E^N_d=1.7$ eV and $E^{NV^-}_d=2.7-2.8$ eV, respectively \cite{Deak}. The positively charged NV center, NV$^+$, acts as an acceptor with ionization energy $E^{NV^+}_a= 0.9-1.1$ eV (see Fig.~\ref{fig_bandgap}) \cite{Deak}. We take the areal density of implanted nitrogens to be $D_s=10^{12}$ cm$^{-2}$. Thus, the volume density of substitutional nitrogen and NV are given by $N^N_d=0.96 D_s/d_{max}$ and $N^{NV}_a=0.04 D_s/d_{max}$, where we take $d_{max}=14$ nm as the maximum implantation depth of nitrogen ions.

The charge density inside the diamond is given by the density of electrons $n(z)$, density of holes $p(z)$, and density of ionized donors and ionized acceptors
\begin{equation}
\rho_d (z)=e\left[{p(z)-n(z)+N^{+}_d(z)-N^{-}_a(z)}\right],
\label{eq:dens1}
\end{equation}
where $e > 0$ is the magnitude of the electron charge. 
Here, we have used the subscript $d$ for the charge density to avoid confusion with the charge density inside the electrolyte solution. Assuming that we can use the Boltzmann approximation, the density of electrons and holes, in the presence of a potential $\phi(z)$, can be written as \cite{Ashcroft}
\begin{equation}
n(z)=N_c \exp \left[{\frac{\mu_0+e \phi(z)-E_c}{k T}}\right], \qquad p(z)=N_v \exp \left[{\frac{E_v-\mu_0-e \phi(z)}{k T}}\right].
\end{equation}
Here, $k$ is the Boltzmann constant, $T$ is the temperature, $N_c$ and $N_v$ are the effective density of states in the conduction and valence bands, respectively, given as
\begin{equation}
N_c=2\left({\frac{m^*_n k T}{2 \pi \hbar^2}}\right)^{3/2}, \qquad N_v=2\left({\frac{m^*_p k T}{2 \pi \hbar^2}}\right)^{3/2},
\end{equation}
and 
\begin{equation}
\mu_0=\frac{E_v+E_c}{2}+\frac{3}{4} k T \ln\left({\frac{m^*_p}{m^*_n}}\right),
\end{equation}
in which $\hbar$ is the reduced Planck constant, $E_v$ is the valence band maximum, $E_c$ is the conduction band minimum, $m^*_n=0.57 m_0$ and $m^*_p=0.8 m_0$ are the effective mass of conduction and valence bands, respectively\cite{Deak}.  The corresponding densities of ionized donors and acceptors are given by 
\begin{equation}
N^{+}_d(z)=\frac{N_d}{1+ \exp\left[{\frac{\mu_0+e\phi(z)-E_d}{k T}}\right]},\quad N^{-}_a(z)=\frac{N_a}{1+ \exp\left[{\frac{E_a-\mu_0-e\phi(z)}{k T}}\right]}.
\end{equation}
\begin{figure}[b!]
\centering
	\includegraphics[width=0.95\textwidth]{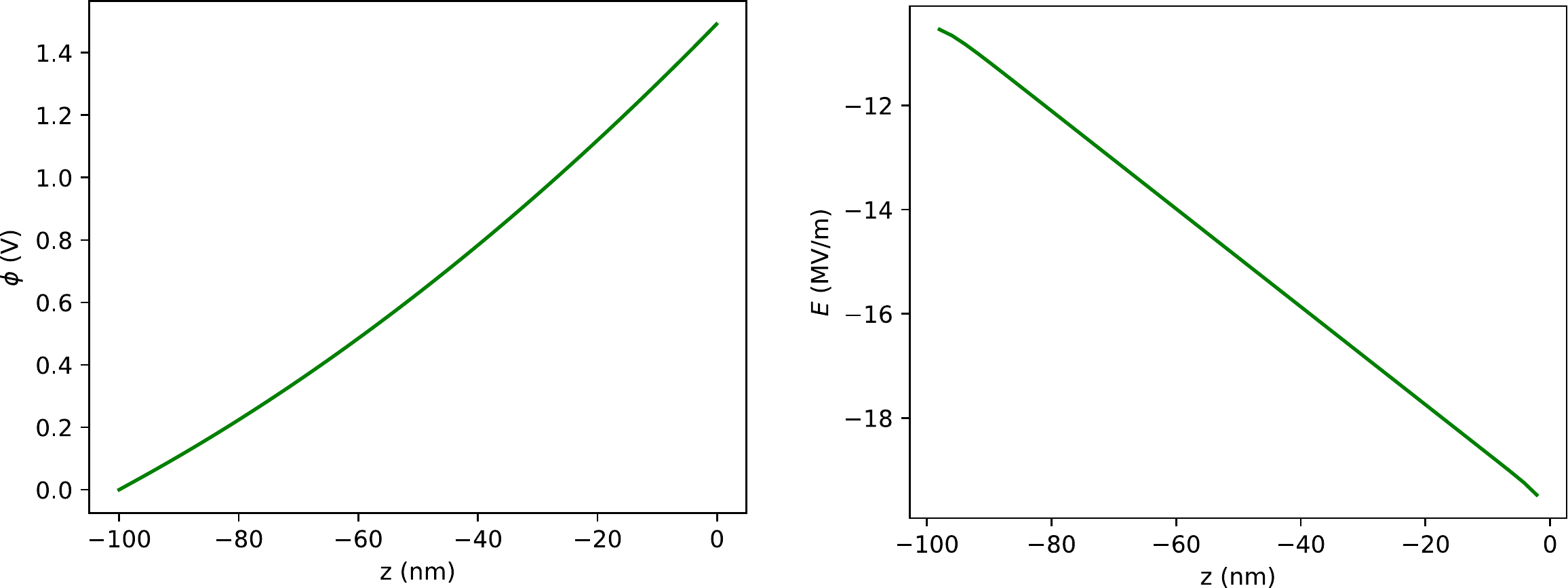}
	\caption{Electric potential, $\phi$, and electric field, $E$, inside the diamond as a function of $z$, depth in the diamond, calculated from the system of equations Eq.~\eqref{eq:dens1}--Eq.~\eqref{eq_phi0_z}. The parameters are $\phi_{be}=1.5$ V, $\phi_{bd}=0$ V, $A=4~{\rm mm^2}$, $\Delta=1$ mm, $c_b=1~{\rm mol/m^3}$, $z_s=2$, $\mathcal{D}=2.3 \times 10^{-9}~{\rm m^2/s}$, and $T=298$ K.}
\label{fig_phi_Ed}
\end{figure}

To find the  potential $\phi(z)$ inside the diamond we use the Poisson equation \begin{equation}
\frac{d^2\phi(z)}{dz^2}=-\frac{\rho_d (\phi)}{\epsilon_d},
\end{equation}
where $\epsilon_d=5.8 \epsilon_0$ is the dielectric constant of diamond with $\epsilon_0$ being the vacuum permittivity. At the interface of electrolyte and diamond we impose the continuity of the electric potential and displacement, i.e.,
\begin{eqnarray}
&&\phi(z=0)|_d=\phi(z=0)|_e=\phi(0), \\
&&\quad\quad\epsilon_d \frac{d\phi}{dz}|_d=\epsilon_e \frac{d\phi}{dz}|_e.
\end{eqnarray}
Here, $d$ and $e$ stands for diamond and electrolyte, and $\epsilon_e$ is the dielectric constant of electrolyte.
Solving the Poisson equation, we obtain the electric field as
\begin{equation}\label{dphidz_d}
E(z)=-\frac{d\phi}{dz}={\rm sgn} (V_0) \sqrt{\left(\frac{\epsilon_e}{\epsilon_d}E_e (0)\right)^2-\frac{2}{\epsilon_d}\int_{{\phi(0)}}^{\phi(z)}{\rho_d(\phi) d\phi}}.
\end{equation}
 Here, we have assumed that $\rho_d$ depends on $\phi$ and does not depend explicitly on $z$. Integrating the above equation, gives the potential $\phi$ at the position $z$ as
\begin{equation}\label{eq_phi0_z}
\int_{{\phi(0)}}^{\phi(z)}{\frac{d\phi}{ \sqrt{\left(\frac{\epsilon_e}{\epsilon_d}E_e (0)\right)^2-\frac{2}{\epsilon_d}\int_{{\phi(0)}}^{\phi}{\rho_d(\phi) d\phi}}}}=-{\rm sgn} (V_0)\int_{0}^{z}{dz}.
\end{equation}

The above equation has to be solved numerically for a specific value of $z$. Replacing the obtained potential in Eq.~\eqref{dphidz_d}, we obtain the electric field at the position $z$. The electric field and potential inside the bulk diamond is shown in Fig.~\ref{fig_phi_Ed} for $c_b=1$ mol/m$^3$. In this figure, the potential in the bulk of electrolyte is taken to be $\phi_{b,e}=1.5$ V at $\Delta=1$ mm. The potential on the surface of the diamond is then found by solving numerically Eq.~\eqref{eq_phi0_z}, by setting the potential inside the bulk of the diamond, at $100$ nm depth, to $\phi_{bd}=0$ V. Note that, we have taken the direction of $z$ axis towards the bulk of the electrolyte. The other parameters are taken as $T=298$ K, $A=4$ mm$^2$, $\mathcal{D}=2.3\times 10^{-9}$ m$^2$/s.

\begin{figure}[t!]
\centering
	\includegraphics[width=0.75\textwidth]{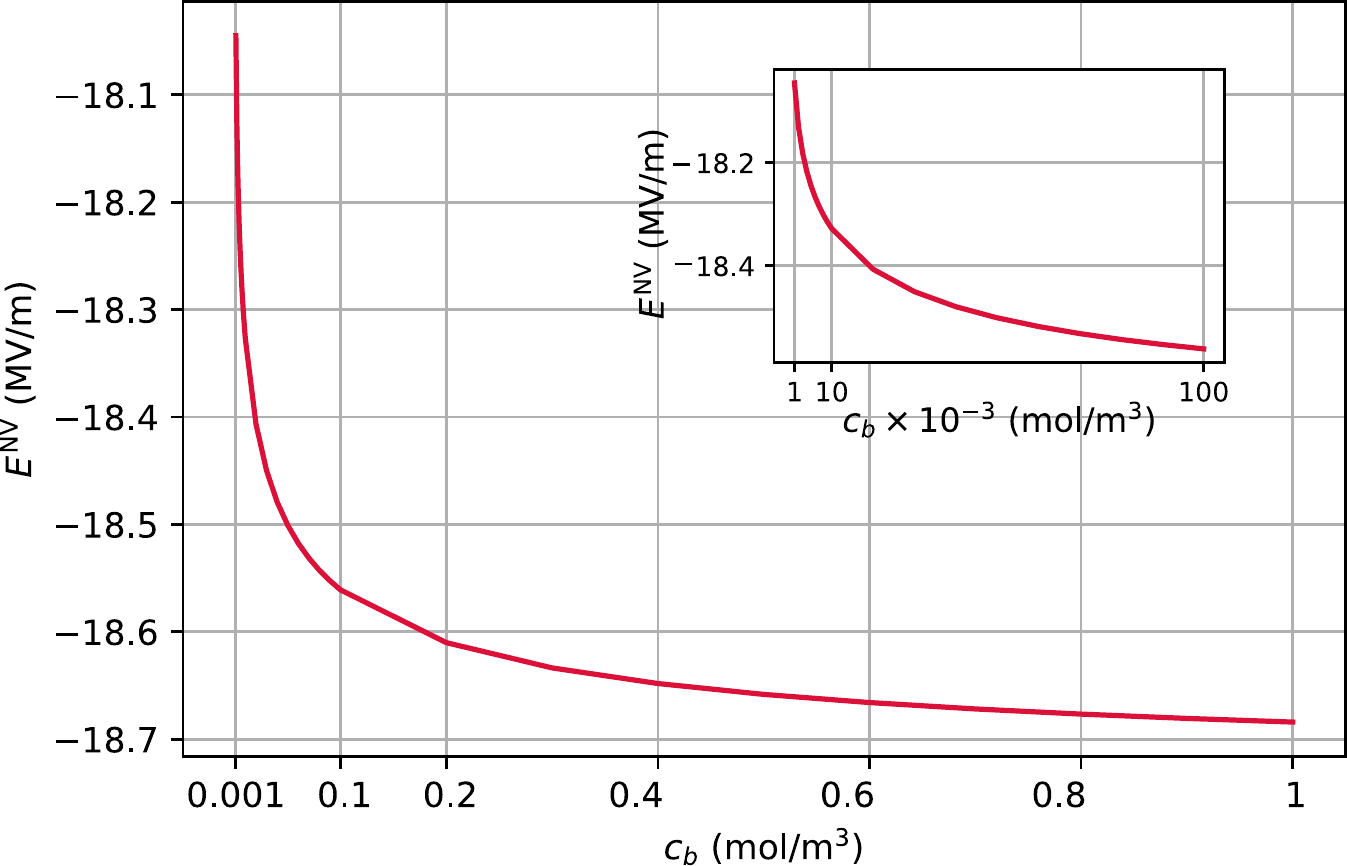}
	\caption{Electric field at the position of NV at 10 nm versus concentration of ions in the bulk $c_b$, calculated from the system of equations Eq.~\eqref{eq:dens1}--Eq.~\eqref{eq_phi0_z}. The parameters are the same as in Fig.~\ref{fig_phi_Ed}. For the range of $c_b$ plotted, we have $\kappa \Delta \gg 1$. The inset shows the electric field for a smaller range of $c_b$.}
\label{fig_Env}
\end{figure}

Figure~\ref{fig_Env} shows the electric field at the position of NV, at 10 nm, as a function of $c_b$, calculated from the system of equations Eq.~\eqref{eq:dens1}--Eq.~\eqref{eq_phi0_z}. The inset of this figure shows that for values of $c_b<0.1~{\rm mol/m^3}$ the gradient of the electric field is higher. In section \ref{section_Esensing}, we will show that this feature may be used to sense the concentration of ions in the electrolyte solution through the Stark effect on the electron spin of the NV center.

\subsubsection{NV center inhomogenous dephasing rate}\label{sec_NV_T2s}
The electron spin of the NV center has a spin triplet ground state. Its ground state Hamiltonian in the presence of a magnetic and an electric field is given by \cite{Ariel}
\begin{equation}\label{eq_Hamiltonian}
H=D S^2_z+\gamma_e \vec{B}\cdot\vec{S}+H_{E_0}+H_{E_1}+H_{E_2},
\end{equation}
where $D=2.87$ GHz is the ground state zero field splitting, $\gamma_e=2.8$ MHz/G is the electron gyromagnetic ratio, and $H_{E_i}$ are the terms due to the electric field that cause transitions between the spin states with the difference in the spin projection $\Delta m_s=i$,
\begin{eqnarray}
&&H_{E_0}=d_{||}S^2_z E^{\rm NV}_z,\\
&&H_{E_1}=d'_{\perp}\left[{E^{\rm NV}_x \left\{S_x,S_z\right\}+E^{\rm NV}_y\left\{S_y,S_z\right\}}\right], \\
&&H_{E_2}=d_{\perp}\left[{E^{\rm NV}_x\left(S^2_y-S^2_x\right)+E^{\rm NV}_y\left\{S_x,S_y\right\}}\right]. \label{eq_HE2}
\end{eqnarray}
Here, $E^{NV}_i$ are the components of the electric field in the NV reference frame. The coupling parameters are experimentally found to be $d_{||}=0.35$ Hz cm/V and $d_{\perp}=17$ Hz cm/V \cite{strain}. On the other hand, the coefficient $d'_{\perp}$ is expected to be of the same order of magnitude as $d_{\perp}$ \cite{DohertyPRB}.
We assume that the NV symmetry axis sits in the $x,z$ plane of the laboratory frame. Therefore, for an electric field which is in the $z$ direction of the lab frame, in the NV frame we have
\begin{equation}
E^{NV}_x=\sqrt{\frac{2}{3}}E_z, \quad E^{NV}_y=0,\quad E^{NV}_z=\sqrt{\frac{1}{3}}E_z.
\end{equation}
 
Due to the large value of the zero field splitting $D$ and the small value of $d_{||}$ and $d_{\perp}$ we can neglect $H_{E_0}$ and $H_{E_1}$ terms in the Hamiltonian. Moreover, for weak magnetic fields, $B\ll D/\gamma$, we can neglect the perpendicular component of the magnetic field. Therefore, the eigenstates of the Hamiltonian in terms of the eigenstates of $S_z$ are given by $\left|0\right\rangle$ and \cite{DohertyPRB}
\begin{eqnarray}
&&\left|  +  \right\rangle =\cos(\theta/2) \left|  +1\right\rangle + \sin(\theta/2) e^{i\varphi_E} \left|-1\right\rangle, \\
&&\left|  -  \right\rangle =\sin(\theta/2) \left|  +1\right\rangle - \cos(\theta/2) e^{i\varphi_E} \left|-1\right\rangle.
\end{eqnarray}
Here, $\tan \theta=\xi_{\perp}/\beta_z$, $\tan \varphi_E=E^{\rm NV}_y/E^{\rm NV}_x$, $\xi_{\perp}=d_{\perp}\sqrt{(E^{\rm NV}_x)^2+(E^{\rm NV}_y)^2}$, and $\beta_z=\gamma B_z$. The energy splitting between $\left|0\right\rangle$ and $\left|\pm\right\rangle$ states is given by 
\begin{equation}
\nu_{\pm}=D\pm\sqrt{\xi_{\perp}^2+\beta^2_z}.
\end{equation}
With our choice of the NV frame, we have $E^{\rm NV}_y=0$ and therefore $\xi_{\perp}=d_{\perp}E^{\rm NV}_x$. It is shown in Ref.~\cite{Jamonneau} that for nonzero values of $B_z$ the electron spin is protected from electric field noise. Therefore, to be able to detect electric field fluctuations we consider $B_z=0$.
For the case of $B_z=0$, the fluctuations in $\nu$ can be written as
\begin{equation}
\delta\nu =-\xi_{\perp}+\sqrt{(\delta \beta_z)^2+(\xi_{\perp}+\delta \xi_{\perp})^2}.
\end{equation}
Assuming that $\delta \xi_{\perp}, \delta \beta_z << \xi_{\perp}$, we can expand the square root and obtain
\begin{equation}
\delta \nu=\delta \xi_{\perp}+\frac{\left({\delta \beta_z}\right)^2}{2\xi_{\perp}}.
\end{equation}
Therefore, 
\begin{equation}\label{correlator_nu}
\left\langle{\delta \nu (t)\delta\nu (t')}\right\rangle=\left\langle{\delta \xi_{\perp} (t)\delta\xi_{\perp} (t')}\right\rangle+\frac{1}{4\xi^2_{\perp}}\left\langle{(\delta \beta_z )^2(\delta\beta_z )^2}\right\rangle,
\end{equation}
where we have assumed $<(\delta \beta_z)^2\delta\xi_{\perp}>=0$.
\begin{figure}[b!]
\centering
	\includegraphics[width=0.5\textwidth]{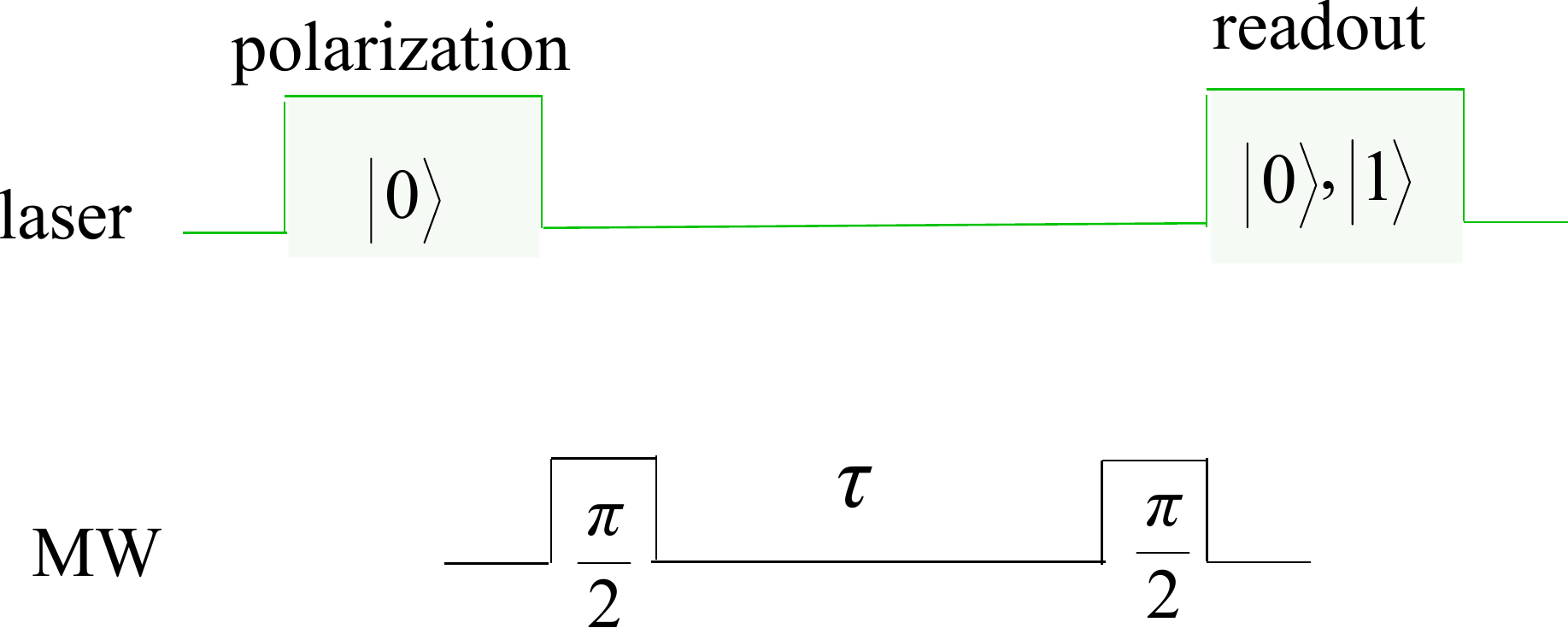}
	\caption{The sequence for free induction decay measurement. The electron spin is initially prepared in $|0\rangle$ using a green laser. With the application of a $\pi/2$ microwave (MW) pulse the electron is prepared in the superposition $\frac{1}{\sqrt{2}}(|0\rangle+|1\rangle)$. The electron spin then goes under a free evolution for time $\tau$. The accumulated phase during the free evolution is projected to the populations using another $\pi/2$ MW pulse. At the end the state of the electron is read out using a green laser.}
\label{fig_NV_pulse}
\end{figure}

The fluctuations on the energy splitting $\nu$ result in dephasing of the electron spin of the NV center which can be measured through free induction decay measurement \cite{JeroNJP}. For this purpose, the spin is prepared in $\left|0\right\rangle$ followed by a  Ramsey sequence which consists of a $\pi/2$ microwave pulse, a free evolution for time $\tau$, and another $\pi/2$ microwave pulse followed by a projective measurement on the electron spin (See Fig.~\ref{fig_NV_pulse}). The probability to detect the electron spin in the state $\left|0\right\rangle$ after the Ramsey sequence is given by
\begin{equation}
P_0(\tau)=\frac{1}{2}\left[1-\cos(\psi+\delta \psi)\right],
\end{equation}
where
\begin{equation}\label{eq_phase}
\psi=\int_{0}^{\tau}{2\pi (\nu_+-\nu_{\rm mw}) dt}, \quad \delta \psi=\int_{0}^{\tau}{2\pi (\delta \nu) dt}.
\end{equation}
Averaging the probability $P_0(\tau)$ and assuming that $\delta \psi$ is normally distributed, we obtain
\begin{equation}\label{FID_signal}
\left\langle P_0(\tau)\right\rangle=\frac{1}{2}\left[1-e^{-\left\langle{\delta \psi^2}\right\rangle/2}\cos(\psi)\right].
\end{equation}
The exponential decay factor determines the dephasing rate of the NV center. To calculate $\left\langle{\delta \psi^2}\right\rangle$ we write
\begin{equation}
\left\langle{\delta \psi^2}\right\rangle=4\pi^2\int_{0}^{\tau}{dt \int_{0}^{\tau}{dt'}\left\langle{\delta\nu(t) \delta\nu(t') }\right\rangle}.
\end{equation}
We assume that the fluctuations are stationary, i.e, its correlation is a function of the time difference, $\left\langle{\delta\nu(t_1)\delta\nu(t_2)}\right\rangle=f(t_1-t_2)$. We define $\tau=t-t'$ and $T=(t+t')/2$. Thus, 
\begin{equation}
\int_{0}^{\tau}{dt \int_{0}^{\tau}{dt'}\left\langle{\delta\nu(t) \delta\nu(t') }\right\rangle}=\int_{0}^{\tau}{d T\int_{}^{}{d \tau \left\langle{\delta\nu(T+\tau/2)\delta\nu(T-\tau/2)}\right\rangle}}.
\end{equation} 
From Eq.~\eqref{correlator_nu}, the correlation function of the fluctuations in $\nu$ due to the electric field can be written 
\begin{equation}
\left\langle{\delta \nu (\tau)\delta\nu (0)}\right\rangle=\left\langle{\delta \xi_{\perp} (\tau)\delta\xi_{\perp} (0)}\right\rangle=d^2_{\perp}\left\langle{\delta E^{\rm NV}_{x} (\tau)\delta E^{\rm NV}_{x} (0)}\right\rangle.
\end{equation}

The electric field at the position of the NV center depends on the electric field at the interface between the electrolyte liquid solution and the diamond, i.e. at $z = 0^+$ in our coordinate system (see Fig.~\ref{fig:geometry}). Therefore, the fluctuations of the electric field at the position of the NV center can be written as
\begin{equation}
\delta E_x^{\rm NV}=\frac{\partial E_x^{\rm NV}}{\partial E_{e}(z=0)} \delta E_{e}(z=0).
\end{equation}
As a result
\begin{equation}\label{eq_correlation_NV}
\left\langle{\delta E_x^{\rm NV}(t){\delta E_x^{\rm NV}(t=0)}}\right\rangle=\left(\frac{\partial E_x^{\rm NV}}{\partial E_{e}(z=0)}\right)^2 \left\langle {\delta E_{e}(z=0,t)\delta E_{e}(z=0,t=0)} \right\rangle.
\end{equation}
We numerically calculate the partial derivative of $E^{\rm NV}$ with respect to $E_e(0)$ using Eq.~\eqref{dphidz_d}. Note that, $\phi$, that appears as the upper limit of the integral, is also a function of $E_e$. For the correlation of the fluctuations of the electric field on the surface, we use Eq.~\eqref{correlator_Ee_simplified}. For $\Delta \approx 1$ mm and for times up to a few miliseconds, we have $\Delta/\sqrt{4 \mathcal{D} t}>>1$. Therefore, the error functions can be approximated by 1 and the second term in the summation can be approximated by zero, i.e.,
\begin{equation}
\langle \delta E_e(0,t) \delta E_e(0,0) \rangle \approx   \frac{\mathcal{F}^2 z_s^2 \Delta c_b}{N_A A \epsilon_e^2}.
\end{equation}
 As a result, the exponential decay factor in the free induction decay signal, Eq.~\eqref{FID_signal}, scales as $t^2$. The factor of $t^2$ in the exponential gives $1/(T^\star_2)^2$, i.e.
\begin{equation}
\left(T^{\star}_2\right)^{-2} = \frac{1}{2}\left.\frac{\partial^2}{\partial t^2}\langle\delta\psi^2\rangle\right|_{t = 0}.
\label{eq:T2}
\end{equation}

\begin{figure}[t!]
\centering
\includegraphics[width=0.75\textwidth]{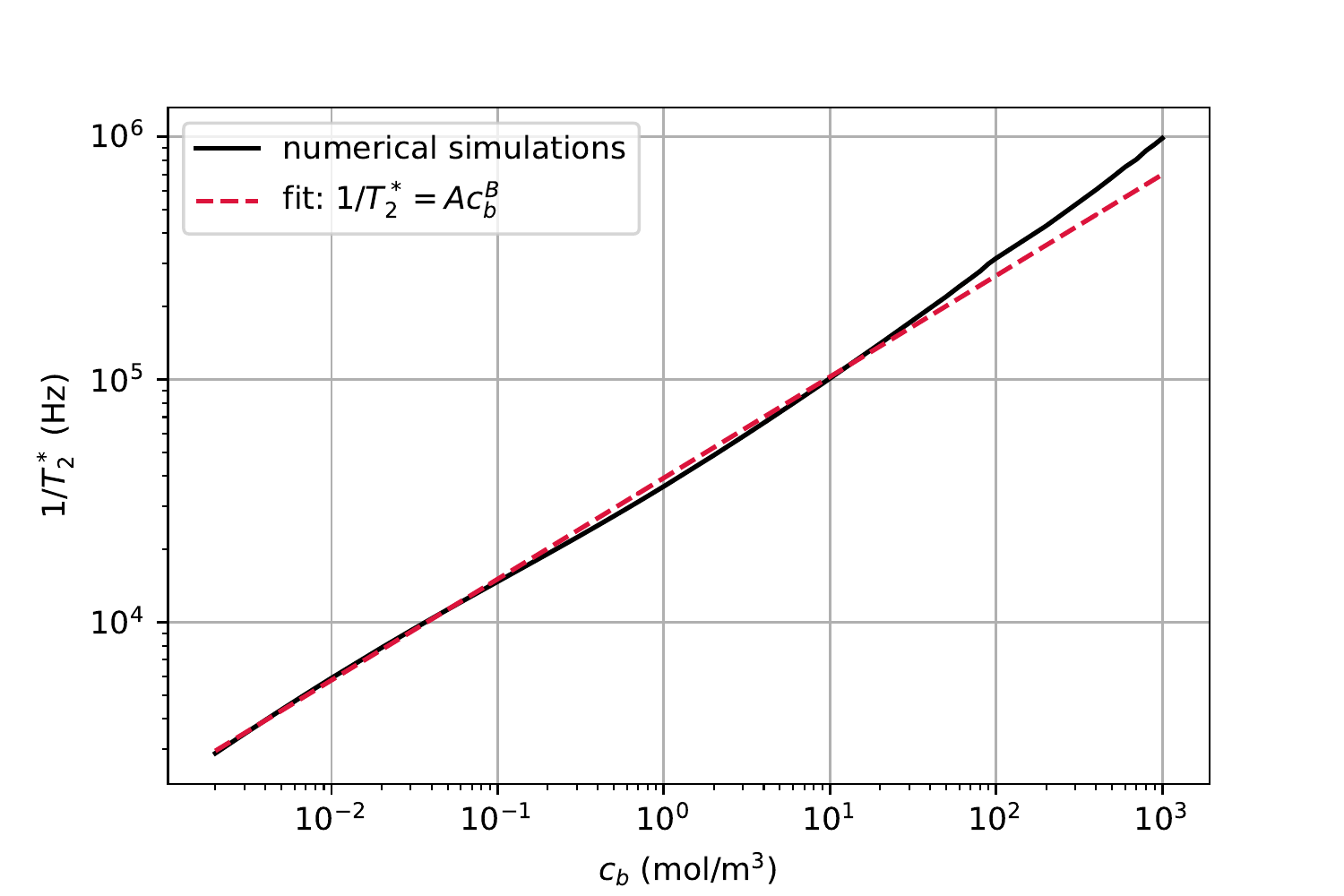}
\caption{ $1/T^\star_2$ for the electron spin of an NV center at the depth of 10 nm as a function of the concentration in the bulk, $c_b$, calculated from Eq~\eqref{eq:T2} (solid black line), and the fit (dashed red line) given by $1/T^\star_2= A c^B_b$ with $A\approx 39295$ Hz (mol/m$^3$)$^{-B}$ and $B\approx 0.417$ (unitless). We have set the potential inside the bulk of the electrolyte to $\phi_{be}=1.5$ V and the potential inside the bulk of diamond to $\phi_{bd}=0$ V. The other parameters are the same as in Fig.~\ref{fig_phi_Ed}.}
\label{T2sfig}
\end{figure}

In Fig.~\ref{T2sfig} we have plotted $1/T^\star_2$ as a function of $c_b$, for an NV center at a depth of 10 nm in the diamond. For the numerical simulations (solid line) we have considered $\phi_{b,e}=1.5$ V and $\phi_{b,d}=0$ V. The solid line is obtained by finding the potential on the surface of the diamond numerically using Eq.~\eqref{eq_phi0_z}. The dashed red line is a fit to the simulations. The fit function is $1/T^\star_2=A c^B_b$ with $A\approx 39295$ Hz (mol/m$^3$)$^{-B}$ and $B\approx 0.417$ (unitless). In Appendix \ref{AppendixC} we derive the sensitivity of this scheme in estimating $c_b$. 

For an NV center in a bulk diamond with an abundance of $^{13}$C isotopes ranging from very low to natural, $1/T^\star_2$ due to surrounding nuclear spins is of the order of a few kilohertz to a few hundred kHz, respectively \cite{Hui,Ishi,Maurer}. To be able to detect fluctuations of the concentration, the corresponding $1/T^\star_2$ should be larger than the intrinsic $1/T^\star_2$ of the NV center. For $c_b>100$ mol/m$^3$ the electric field fluctuations produced by the diffusional fluctuations in the local concentration of ions result in $1/T^\star_2>300$ kHz and for $c_b>0.04$ mol/m$^3$ result in $1/T^\star_2> 10$ kHz. Therefore, the range of concentrations that can be estimated with this method depends on the diamond sample properties. We note that, for an NV that is positioned at a smaller distance from the surface of the diamond, the induced electric field at the position of NV is larger and the induced $1/T^{\star}_2$ is larger. 

In the next section, we show that for a suitable range of concentrations the Stark effect may also be used to estimate $c_b$.

\subsubsection{Electric field sensing}\label{section_Esensing}
To be able to measure the electric field, the electron spin needs to be prepared in a state that is susceptible to the term $H_{E_2}$ of the Hamiltonian, Eq.~\eqref{eq_HE2} \cite{Dolde,Michl}. As $H_{E_2}$ term causes transition between $m_s=\pm 1$ state, such a state is a superposition of $|m_s=\pm 1\rangle$ states. A perpendicular magnetic field much smaller than the zero field splitting, $D$, in the absence of an axial magnetic field, can prepare the states \cite{Michl}  
\begin{eqnarray}
|-\rangle=\frac{1}{\sqrt{2}}\left(|+1\rangle-e^{2 i\varphi_B}|-1\rangle\right), \qquad |+\rangle=\frac{1}{\sqrt{2}}\left(|+1\rangle+e^{2 i\varphi_B}|-1\rangle\right),
\end{eqnarray}
where $\varphi_B$ is the azimuthal angle of the magnetic field. The energy shift of these states due to $H_{E_2}$ are
\begin{equation}
\langle-|H_{E_2}|-\rangle=d_{\perp} E^{\rm NV}_{x} \cos(2\varphi_B), \qquad \langle+|H_{E_2}|+\rangle=-d_{\perp} E^{\rm NV}_{x} \cos(2\varphi_B),
\end{equation}
where we have taken $E^{\rm NV}_y=0$. As can be seen from Fig.~\ref{fig_Env}, for $c_b<0.1~{\rm (mol/m^3)}$, a change of one order of magnitude in $c_b$ results in a change of about $230$ kV/m in the electric field at the position of the NV center, $E^{\rm NV}_x$. This change of the electric field results in an energy shift of about 39 kHz for $\varphi_B=0$. This change of the electric field can be resolved in a Ramsey sequence, by preparing the NV center in the superposition state of $|0\rangle$ and either of $|\pm\rangle$ states and measuring the energy shift. 

\section{Conclusions}
We considered an NV center in a bulk diamond in contact with an electrolyte. The fluctuations in the concentration of ions induces a fluctuating electric field at the position of the NV center. We first showed that the dephasing rate of the electron spin of the NV center, $1/T^\star_2$, can be used to estimate the concentration of ions in the electrolyte for a range of concentrations $c_b > 0.04~ {\rm mol/m^3}$ depending on the intrinsic $1/T^{\star}_2$ of the electron spin of the NV. For this range of $c_b$, the induced $1/T^\star_2$ resulting from the fluctuations of the concentration of charged species is larger than 10 kHz. We also showed that for $c_b<0.1$ mol/m$^3$ the gradient of the electric field induced at the position of the NV center is large enough to be resolved through Ramsey spectroscopy for a change of one order of magnitude in $c_b$. Therefore, through the estimation of the electric field, $c_b$ can be estimated.

Using a dynamical decoupling sequence such as spin echo or Carr-Purcell-Meiboom-Gill sequence the $1/T^\star_2$ of the electron spin of the NV center due to surrounding nuclear spins can be increased up to kHz. In that case, the NV center would be sensitive to the fluctuations of the concentration in a wider range of $c_b$.


\vspace{6pt} 



\section{Acknowledgements}
H.T.D.~acknowledges support from the Fondecyt-postdoctorado grant No.~3170922 and Universidad Mayor for a postdoctoral fellowship. E. M. and  J. R. M. acknowledge support from the Fondef Project Grant No.~ID16I10214, and from the Project Grant ANID PIA Anillo ACT192023. E. M.~acknowledges support from Fondecyt Regular Grant No 192023. J.R.M.~acknowledges support from Fondecyt Regular Grant No 1180673, and Air Force grant number FA9550-18-1-0513.
\section{Author contributions}
E. M. and J. R. M. conceptualized the idea. E. M. and H. T. D. performed the calculations and wrote the manuscript. E. M. and J. R. M. supervised the work. J. R. M. reviewed and edited the manuscript. All authors have read and agreed to the published version of the manuscript.






\appendix
\section{Electric potential and electric field inside the electrolyte}\label{appendixA}
In this appendix we give details of the derivation of the electric potential and the electric field inside the electrolyte. First, we define a dimensionless potential as
\begin{equation}
\Phi(z) = \frac{z_s\mathcal{F}}{RT}\left(\phi(z)-\phi(\Delta)\right),
\end{equation}
where $\phi(\Delta)$ is the potential in the bulk of the electrolyte (see Fig.~\ref{fig:geometry}). In the bulk, the solution should be electrically neutral due to charge conservation, i.e.,
 \begin{equation}
 \sum_{s=\pm}{z_s c_{b,s}}=0,
 \end{equation}
with $z_s$ being the valence of the ion species and $c_{b,s}$ being the bulk concentration of species. Assuming $z_+=-z_-=z_s$ we can write $c_{b,+}=c_{b,-}=c_b$. 
Therefore, Poisson equation, given in Eq.~\eqref{eq:concentration-diff-zero3}, in terms of the dimensionless potential can be written as
\begin{equation}
\label{eq:poisson2}
\frac{\partial^2 \Phi^{}}{\partial z^2} = 2\frac{z^2_s \mathcal{F}^2 c_b}{RT\epsilon_e}\sinh{\left(\Phi^{}(z)\right)}.
\end{equation}

We first consider the limit where $c_b \to 0$. Since $\Phi$ is proportional to $c_b$, because the second derivative of $\Phi$ is proportional to $c_b$, in the limit of $c_b \to 0$, we can linearize the above equation and obtain  
\begin{equation}
\frac{\partial^2 \Phi^{}}{\partial z^2} = \kappa^2 \Phi(z),
\end{equation}
where $\kappa$ is defined in Eq.~\eqref{eq_kappa}. The solution of this differential equation can be written as 
\begin{equation}
\Phi(z)=A \cosh(\kappa (\Delta -z))+B \sinh(\kappa (\Delta -z)).
\end{equation}
The constants $A$ and $B$ can be obtained by considering the boundary conditions $\Phi(0)=z_s\mathcal{F}V_0/(R T)$ and $\Phi(\Delta)=0$. Thus, we obtain
\begin{equation}
\Phi(z)=\frac{\Phi(0)\sinh(\kappa(\Delta-z))}{\sinh(\kappa \Delta)}.
\end{equation}
Therefore, the electric field can be written as
\begin{equation}
    E(z)=-\frac{d\phi}{dz}=\frac{\kappa V_0 \cosh(\kappa(\Delta-z))}{\sinh(\kappa \Delta)}.
\end{equation}
In the limit of $c_b\to 0$, i.e., $\kappa \to 0$ we obtain 
\begin{equation}
    E(z)=\frac{\kappa V_0}{\kappa \Delta}=\frac{V_0}{\Delta}.
\end{equation}
This is the expected electric field between two parallel plates with the distance $\Delta$ and the potential difference $V_0$. 

We now obtain the exact solution of Eq.~\eqref{eq:poisson2}. We first define the dimensionless quantity $\xi = \kappa z$ where $\kappa$ is the inverse screening length given in Eq.~\eqref{eq_kappa}. 
Therefore, we can write
\begin{equation}
\frac{\partial^2\Phi^{}(\xi)}{\partial \xi^2}= \sinh{\left(\Phi^{}(\xi)\right)}.
\end{equation}
Multiplying by  $\partial\Phi^{}/\partial \xi$ and integrating over the interval $[\xi, \kappa \Delta]$ we obtain
 \begin{equation}
\left(\frac{\partial\Phi^{}}{\partial \xi}\right)^2_{\kappa \Delta}  - \left(\frac{\partial\Phi^{}}{\partial \xi}\right)^2_{\xi}   = 2[\cosh(\Phi(\kappa \Delta))-\cosh(\Phi(\xi))].
\end{equation}
From Eq.~\eqref{eq:poisson2} we have $\Phi^{}(\kappa \Delta) = 0$. Moreover, for $\kappa \Delta \gg 1$, due to ion screening we can assume $\partial\Phi^{}/\partial\xi|_{\kappa \Delta}=0$. Thus, we can write
\begin{equation}
\left(\frac{\partial\Phi^{}}{\partial \xi}\right)^2   = 2(\cosh(\Phi(\xi))-1)= 4 \sinh^2(\Phi(\xi)/2).
\end{equation}
In Section \ref{sec_diffusion}, we introduced $\phi^{}(0)-\phi(\Delta) = {V}_0$. Therefore, if ${V}_0>0~(<0)$ the slope of $\Phi^{}$ must be negative (positive). As a result, we can write
\begin{equation}\label{eq:phiprime}
\partial\Phi^{}/\partial \xi = -2 \sinh(\Phi(\xi)/2).
\end{equation}
 This is a separable equation which can be integrated directly, yielding
\begin{equation}
 \frac{1-e^{-\Phi^{}/2}}{1+e^{-\Phi^{}/2}}= \tanh\left({\frac{z_s \mathcal{F}{V}_0}{4RT}}\right) e^{-\xi}.
\end{equation}
Solving for $\Phi^{}$ gives
\begin{equation}\label{eq:pot0}
\Phi^{}(\xi) =  2 \ln\left[\frac{1+\tanh(\Phi(0)/4)e^{-\xi}}{1-\tanh(\Phi(0)/4)e^{-\xi}}\right]
\end{equation}
Therefore, in terms of $\phi$ and $z$ we can write 
\begin{equation}\label{eq_phie_appendix}
\phi_{}(z) - \phi(\Delta)= \frac{2 R T}{z_s \mathcal{F}}\ln\left[\frac{1+\tanh(\frac{z_s \mathcal{F} V_0}{4 R T})\exp({-\kappa z })}{1-\tanh(\frac{z_s \mathcal{F} V_0}{4 R T})\exp({-\kappa z })}\right].
\end{equation}

\section{Thermal fluctuations in the local concentration of chemical species}\label{AppendixB}
In this appendix we derive the fluctuations of the concentration of ions in the electrolyte. From the statistical point of view, the local concentration of chemical species is the average number of particles per unit
volume $\langle n(\mathbf{x},t) \rangle$, where $n(\mathbf{x},t)$ is a random variable subject to
stochastic dynamics due to Brownian motion. The corresponding average then satisfies the diffusion equation
\begin{eqnarray}
\frac{\partial}{\partial t}\langle n(\mathbf{x},t) \rangle = \mathcal{D}\nabla^2 \langle n(\mathbf{x},t) \rangle,
\label{eq_dif1}
\end{eqnarray}
where $\mathcal{D}$ is the diffusion constant. This equation possesses a Green's function $G(\mathbf{x},t;\mathbf{x}_0,t_0)$,
that represents the transition probability for a particle to diffuse from a point $\mathbf{x}_0$ at time $t_0$ to a point $\mathbf{x}$
at time $t$. This transition probability satisfies the differential equation
\begin{eqnarray}
\frac{\partial}{\partial t}G(\mathbf{x},t;\mathbf{x}_0,t_0) = \mathcal{D}\nabla^2 G(\mathbf{x},t;\mathbf{x}_0,t_0),
\label{eq_green}
\end{eqnarray}
with the initial condition
\begin{eqnarray}
G(\mathbf{x},t_0;\mathbf{x}_0,t_0) = \delta(\mathbf{x} - \mathbf{x}_0).
\end{eqnarray}
Equation \eqref{eq_green} has an explicit solution (in $d$ dimensions) given by
\begin{eqnarray}
G(\mathbf{x},t;\mathbf{x}_0,t_0) = \frac{1}{\left(4 \pi \mathcal{D} (t - t_0) \right)^{d/2}}\exp\left({-\frac{|\mathbf{x} - \mathbf{x}_0|^2}{4 \mathcal{D} (t - t_0)}}\right).
\label{eq_green_sol}
\end{eqnarray}

Following Ref.~\cite{vanKampen}, we define the two-particle correlator
\begin{eqnarray}
[n(\mathbf{x}_1,t)n(\mathbf{x}_2,t)] \equiv \langle n(\mathbf{x}_1,t)n(\mathbf{x}_2,t) \rangle - \langle n(\mathbf{x}_1,t)\rangle \langle n(\mathbf{x}_2,t)\rangle
- \delta(\mathbf{x}_1 - \mathbf{x}_2)\langle n(\mathbf{x}_1,t)\rangle.
\label{eq_corr1}
\end{eqnarray}
It can be shown that this two-particle correlator satisfies the following equation
\begin{eqnarray}
\frac{\partial}{\partial t}[n(\mathbf{x}_1,t)n(\mathbf{x}_2,t)] = \mathcal{D}\left(\nabla^2_1 + \nabla^2_2 \right)[n(\mathbf{x}_1,t)n(\mathbf{x}_2,t)].
\label{eq_corr2}
\end{eqnarray}
We now define the particle number fluctuations $\delta n(\mathbf{x},t) = n(\mathbf{x},t) - \langle n(\mathbf{x},t)\rangle$, that by definition
satisfies the following properties:
\begin{eqnarray}
\langle \delta n(\mathbf{x},t) \rangle &=& 0,\nonumber\\
\langle \delta n(\mathbf{x}_1,t_1) \delta n(\mathbf{x}_2,t_2)\rangle &=&  \langle n(\mathbf{x}_1,t)n(\mathbf{x}_2,t) \rangle - \langle n(\mathbf{x}_1,t)\rangle \langle n(\mathbf{x}_2,t)\rangle \nonumber\\
&=& [n(\mathbf{x}_1,t)n(\mathbf{x}_2,t)]  + \delta(\mathbf{x}_1 - \mathbf{x}_2)\langle n(\mathbf{x}_1,t)\rangle.
\label{eq_corr3}
\end{eqnarray}
Combining Eqs.~\eqref{eq_corr2} and \eqref{eq_corr3}, after some algebraic manipulations, one obtains the differential equation for the correlator of fluctuations in
the particle number
\begin{eqnarray}
\left( \frac{\partial}{\partial t} - \mathcal{D}	\nabla_1^2 - \mathcal{D}\nabla_2^2\right)\langle \delta n(\mathbf{x}_1,t) \delta n(\mathbf{x}_2,t)\rangle = 
2 \mathcal{D} \nabla_1\cdot\nabla_2\left\{ \delta(\mathbf{x}_1 - \mathbf{x}_2)\langle n(\mathbf{x}_1,t)\rangle  \right\}.
\end{eqnarray}
The steady-state (equilibrium) solution to this equation is
\begin{eqnarray}\label{eq_equil_correl}
\langle \delta n(\mathbf{x}_1) \delta n(\mathbf{x}_2)\rangle_{\rm eq} = \langle n(\mathbf{x}_1) \rangle_{\rm eq}\delta(\mathbf{x}_1 - \mathbf{x}_2).
\end{eqnarray}
We can now consider the time evolution of the fluctuation correlator, by using the conditional probability (Green's function) as follows
\begin{eqnarray}
\langle \delta n(\mathbf{x}_2,t_2) \delta n(\mathbf{x}_1,t_1)\rangle = \int d\mathbf{x}_3\, G(\mathbf{x}_2,t_2;\mathbf{x}_3,t_1)\langle \delta n(\mathbf{x}_3) \delta n(\mathbf{x}_1)\rangle_{\rm eq}.
\label{eq_convolution}
\end{eqnarray}
Inserting the Green's function from Eq.~\eqref{eq_green_sol} and calculating the integral, we obtain
\begin{eqnarray}
\langle \delta n(\mathbf{x}_2,t_2) \delta n(\mathbf{x}_1,t_1)\rangle =  \frac{\langle n(\mathbf{x}_1) \rangle_{eq}}{\left(4 \pi \mathcal{D} (t_2 - t_1) \right)^{d/2}}\exp\left({-\frac{|\mathbf{x}_2 - \mathbf{x}_1|^2}{4 \mathcal{D} (t_2 - t_1)}}\right)\Theta(t_2 - t_1).
\end{eqnarray}
Here, $\Theta(t)$ is the step function. This equation clearly shows an exponential decay of the correlation of the fluctuations with the relative distance $|\mathbf{x}_2 - \mathbf{x}_1|$.

For the particular case of a three-dimensional system that, due to physical constraints, is subjected to concentration gradients and fluctuations along only one dimension (for instance $z$-direction), i.e. $n(\mathbf{x},t)=n(z,t)$, the particle
concentration is still defined per unit volume, i.e., in units of $1/m^{3}$. In this case, the equilibrium correlator, Eq.~\eqref{eq_equil_correl}, must be modified as
\begin{eqnarray}
\langle \delta n(z_1) \delta n(z_2)\rangle_{eq} = \frac{\langle n(z_1)\rangle_{eq}}{A} \delta(z_1 - z_2),
\label{eq_equil_correl1d}
\end{eqnarray}
where the Dirac delta function is one-dimensional (with units of $1/m$), and $A$ is the transversal area to the direction of the gradients (the surface area of the bulk diamond).

With this argument, Eq.~\eqref{eq_convolution} is now given by a one-dimensional integral, with the Green's function for $d=1$, as 
\begin{eqnarray}
\langle \delta n(z_1,t) \delta n(z_2,0)\rangle &=& \int dz_3\, G(z_2,t;z_3,0)\langle \delta n(z_3) \delta n(z_1)\rangle_{\rm eq}\nonumber\\
&=& \frac{\langle n(z_1) \rangle_{eq}}{A\left(4 \pi \mathcal{D}\, t \right)^{1/2}}\exp\left[-{\frac{(z_2 - z_1)^2}{4\mathcal{D} t}  }\right]\Theta(t).
\label{eq_convolution_1d}
\end{eqnarray}
From the above equation, the concentration fluctuations on a molar basis, for two species $s,s' = 1,2$, corresponds to
\begin{eqnarray}
\langle \delta c_s(z_1,t) \delta c_{s'}(z_2,0)\rangle = \delta_{s,s'} \frac{c_s^{eq}(z_1)}{N_A A\left(4 \pi \mathcal{D}_s\, t \right)^{1/2}}\exp\left[-{\frac{(z_1 - z_2)^2}{4\mathcal{D}_s t}  }\right]\Theta(t),
\end{eqnarray}
where $N_A = 6.02\times 10^{23}$ (particles/mole) is Avogadro's number, and $\mathcal{D}_s$ the diffusion constant for each species. 
\section{Sensitivity}\label{AppendixC}
In this appendix we derive the sensitivity of the proposed scheme in estimating $c_b$. The sensitivity, $\eta$, is defined as the square root of the product of the variance of $c_b$ and the measurement time
\begin{equation}
    \eta=\sqrt{(\Delta c_b)^2 t}.
\end{equation} 
As $c_b$ is estimated by estimating the dephasing time $T^\star_2$, from the error propagation formula we can write
\begin{equation}
    (\Delta c_b)^2=\frac{(\Delta T^\star_2)^2}{|\partial T^\star_2/\partial c_b|^2}.
\end{equation}
From the fit to the numerical simulations of $1/T^\star_2$ versus $c_b$ (see Fig.~\ref{T2sfig}), we have $\partial T^\star_2/\partial c_b=-(B/A) c^{-B-1}_b$ where $A\approx 39295$ Hz (mol/m$^3$)$^{-B}$ and $B\approx 0.417$ (unitless). On the other hand, $T^\star_2$ is estimated by measuring the electron spin of the NV center through the detection of its photoluminescence. Thus, the observable which is used to estimate $T^\star_2$ can be taken as $M=a|0\rangle\langle0|+b|1\rangle\langle1|$ \cite{Meriles}. The parameters $a$ and $b$ are random variables which determine the number of photons collected during the readout interval (about 300 ns) if the spin state is in $|0\rangle$ and $|1\rangle$ states, respectively, with their averages over many measurements given by $\alpha=\langle a\rangle$ and $\beta=\langle b\rangle$. The variance of $T^\star_2$, can then be written as
\begin{equation}\label{eq_DeltaT2s}
    (\Delta T^\star_2)^2=\frac{(\Delta M)^2}{|\partial\langle M\rangle_{\rm avg}/\partial T^\star_2|^2},
\end{equation}
where $(\Delta M)^2=\langle M^2 \rangle_{\rm avg}-\langle M\rangle^2_{\rm avg}$, with the average being over many measurements.

For a free induction measurement with evolution time $t$ we obtain \cite{Meriles}
\begin{equation}\label{eq_expM}
    \langle M\rangle_{\rm avg}=\frac{\alpha+\beta}{2}+\frac{\alpha-\beta}{2} e^{-(t/T^\star_2)^2} \cos(\psi),
\end{equation}
and 
\begin{equation}\label{eq_DeltaM2}
    (\Delta M)^2=\frac{\alpha+\beta}{2}+\frac{\alpha-\beta}{2} e^{-(t/T^\star_2)^2} \cos(\psi)+\frac{(\alpha-\beta)^2}{4}(1-\cos^2(\psi)e^{-2 (t/T^\star_2)^2}),
\end{equation}
where $\psi$ is the accumulated phase during the evolution time $t$, given in Eq.~\eqref{eq_phase}. Note that, in Section \ref{sec_NV_T2s} we found that the exponential decay factor scales with $t^2$. Taking the derivative of Eq.~\eqref{eq_expM} and substituting into Eq.~\eqref{eq_DeltaT2s} we obtain 
\begin{equation}
    (\Delta T^\star_2)^2=\frac{15 (T^\star_2)^6 e^{2(t/T^\star_2)^2} }{2\alpha~t^4 \cos^2(\psi)},
\end{equation}
where we have used $\alpha=3\beta/2$ and since $\alpha\approx 0.03$, therefore $1/\alpha>>1$, we have only considered the first term in Eq.~\eqref{eq_DeltaM2} \cite{Meriles}. Thus, the sensitivity is obtained as
\begin{equation}
    \eta=\sqrt{\frac{15 t}{2\alpha}}\left(\frac{A}{B}\right)c_b^{B+1}t^{-2}(T^\star_2)^3 e^{(t/T^\star_2)^2}
\end{equation}
where to obtain a limit we have taken $\cos(\psi)\approx 1$. As an example we obtain $\eta$ for $c_b=10$ mol/m$^3$ for which we have found $T^{\star}_2\approx 10~\mu$s. Taking $t=10~\mu$s we obtain $\eta=3.27$ mol m$^{-3}$ Hz$^{-1/2}$. This sensitivity can be improved by a protocol in which $t$ is changed during the measurement repetitions. An extra enhancement may be achieved by optimizing the measurement sequence using a Bayesian approach \cite{HTD}. In addition the sensitivity can also be improved by increasing the photon collection efficiency which can be achieved by for instance coupling the NV to a photonic waveguide \cite{Babinec}, or improving the spin readout of the NV center through spin to charge conversion \cite{Shields}.




\end{document}